\documentstyle[aps,prl,multicol,epsf]{revtex}
\begin{document}
\title{\bf Scaling in a Nonconservative Earthquake Model of Self-Organised 
          Criticality }

\author{Stefano Lise and Maya  Paczuski}
\address{Department of Mathematics, Huxley Building, Imperial College
of Science, Technology, and Medicine, London UK SW7 2BZ \\}
\date{\today}

\maketitle 

\begin{abstract}
We numerically investigate the Olami-Feder-Christensen model for earthquakes 
in order to characterise its scaling behaviour. We show that ordinary finite 
size scaling in the model is violated due to global, system wide events. 
Nevertheless we find that subsystems of linear dimension small compared to 
the overall system size obey finite (subsystem) size scaling, with universal 
critical coefficients, for the earthquake events localised within the 
subsystem.  We provide evidence, moreover, that large earthquakes responsible
for breaking finite size scaling are initiated predominantly near the 
boundary.

\end{abstract}

\vspace{0.3cm}

\begin{multicols}{2}

\section{Introduction}

Many dynamical phenomena in nature are intermittent. This ``bursty'' dynamics 
may be related to an underlying complex state, often characterised by long 
range correlations in space and time. For example, the crust of the earth 
alternates long periods of relative quiescence with burst of activity 
(earthquakes), which have a wide range of possible sizes.  The behaviour 
of earthquakes is described by the empirical Gutenberg-Richter (GR)
law~\cite{gutenberg}, where the distribution of energy dissipated in
earthquake events is a power law over many orders of magnitude in
energy. The GR scaling extends from the smallest measurable earthquakes, 
which are equivalent to a truck passing by, to the most disastrous that 
have been recorded.  Similar scale free behaviour of bursts is observed 
in vastly different kinds of physical systems such as flux motion through 
disordered type II superconductors placed in a magnetic field~\cite{field}, 
or in granular piles, under some conditions~\cite{frette}, etc. 

Self-organised criticality (SOC)~\cite{BTW} has been proposed as a general
dynamical principal behind the observed complex behaviour of many 
natural phenomena.  It refers to the fundamental property of slowly driven,
extended systems to organise, over a sufficiently long transient period, 
into a dynamical critical state which lacks  any characteristic time or length 
scale.  The amplitude of the response of the system to an external 
perturbation follows a power law distribution. A number of simple models have 
been developed to test the applicability of SOC to a variety of complex 
interacting dynamical systems, such as sand piles and earthquakes 
(for a review see e.g. ref.~\cite{bak_book,jen_book,turc_rev}).

One of the basic theoretical problems is to identify robust, and thus
physically relevant mechanisms for SOC to emerge, and to define a
space of parameters and dynamical processes where SOC is a stable
feature.  Although it has been proposed that the presence of
conservation laws (e.g. sand grains being transported in a sand pile)
or special symmetries was necessary for
SOC~\cite{hwa-kardar,grinstein}, many known examples of physical
phenomena and some models have been found where no apparent
conservation law or special symmetry exists. These include, besides
earthquakes, biological evolution, forest fires, epidemics, (possibly)
solar flares,  (possibly) reconnection
events in the magnetotail, etc.~\cite{bak_book,jen_book,turc_rev}.  
In contrast to conservative systems, the mechanisms for SOC in 
non-conservative systems are not very well established.

A nonconservative SOC model that in recent years has attracted much
attention is the so called OFC model~\cite{ofc}. The OFC model is 
a simplified lattice representation of a spring-block model for 
earthquake dynamics which was originally introduced by Burridge and 
Knopoff~\cite{burridge}. The Burridge-Knopoff model can be schematized as 
a two dimensional network of blocks interconnected by springs. All blocks 
are subject to an external driving force, which pulls them, and to a static 
friction, which opposes their motion. In the OFC model, each  site of a 
lattice is associated with a continuous variable, which represent the force 
acting on a block.
A slow driving is applied to the system by increasing uniformly and 
simultaneously the forces of all the elements. When the force on a site 
exceeds some threshold value (the maximal static friction), the site relaxes 
and distributes part of its force to  nearest neighbour sites.   
Each such discharge event is accompanied by a local loss in accumulated force 
from the system. 
This conceptually simple and seemingly numerically tractable model 
reproduces some of the qualitative phenomenology of the statistics of 
earthquake events such as power law behaviour over a range of sizes, and 
intermittency or clustering of large events~\cite{ofc_2}.
Extensions of the model have been recently developed  which reproduce 
to some extent Omori's law and other temporal patterns associated with
earthquakes~\cite{hainzl}.

In the context of nonconservative models, the OFC model is of particular 
interest as it is possible to directly control the level of conservation of 
the dynamics through a parameter $\alpha$. Early analysis on relatively small 
systems indicated that the OFC model exhibited SOC, in the sense that 
earthquakes in the steady state obeyed finite size scaling (FSS) when the 
system size was varied~\cite{ofc}.  However, the critical coefficients 
obtained using the FSS ansatz were found to be nonuniversal. In particular 
the exponents characterising the power law distributions appeared to vary 
with both the dissipation parameter, $\alpha$, and the form of the boundary 
conditions.  This would have been in sharp contrast to the usual fixed
point picture of critical phenomena where most microscopic details are
irrelevant.  Moreover some apparent critical exponents obtained using FSS
violated physical bounds~\cite{klein}, putting some doubt on the existence of
criticality in the model. Recently it was shown using a multiscaling
analysis of large-scale simulations that, actually, the avalanche size
distribution has a universal power law behaviour, independent of the
dissipation parameter and for different boundary conditions, but that
the cutoff in the power-law distribution does not behave according to
FSS~\cite{lisepac}.  In larger systems, proportionally more of the force 
can be dissipated in the largest events that occur, and the cutoff function
becomes sharper and sharper as the system size increases.

Departures from standard FSS have been reported for various SOC models
as, for example, the sandpile model~\cite{stella}, the Drossel-Schwabl (DS)
forest fire model~\cite{schenk,pastor_1} and the Zhang model~\cite{pastor_2}. 
In this paper we address the question of the origin of the breaking of 
FSS and its relation to the mechanism responsible for SOC in the OFC
model. In particular we test the implicit assumption behind the FSS
hypothesis that a finite systems behaves as a subsystem of a larger 
system. The paper is organised as follows. 
In the section II we describe in some detail the model.  In section
III we present the results of our numerical study relative to two different 
type of probability distributions for earthquake sizes. The first distribution
concerns earthquakes which are localised within a given subsystem. We show 
that this subset of earthquakes exhibits ordinary FSS as long as the linear 
extent,  $\lambda$, of the subsystem is sufficiently small compared to the 
linear extent, $L$, of the entire system. The second distribution groups 
earthquakes according to the position of their starting site relative to the 
boundaries of the system. From this investigation, we deduce that FSS is 
violated due to large events initiated in a region near the boundary.
Finally, in section IV we discuss our results and draw some conclusions.

\section{The model}

We consider a two-dimensional square lattice of $L \times L$ sites.  
To each site $i$ of the lattice we associate a continuous variable 
$F_i$, which initially take some random values between zero and a 
threshold value $F_{th}$.  The dynamics proceeds then indefinitely. In the 
limit of infinite time scale separation between the slow driving and the 
(almost) instantaneous earthquake process, the dynamics is:  
\begin{enumerate}
\item {\em Uniform drive}: all forces  $F_i$ are increased at the same rate, 
  until one of them reaches the value $F_{th}$.
\item {\em Earthquake}: when a site becomes unstable (i.e 
  $F_i \geq F_{th}$),  the uniform driving is stopped and the system evolves 
  according to the following local relaxation rule       
     \begin{equation}
     \label{av_dyn}
           F_i \geq F_{th}  \Rightarrow \left\{ \begin{array}{l}
                                       F_i \rightarrow 0 \\
                         F_{nn} \rightarrow F_{nn} + \alpha F_i
                                      \end{array} \right.
      \end{equation} 
   until there are no more unstable sites. In eq.~(\ref{av_dyn}), the subscript
  ``nn'' stands for the four nearest neighbours to site $i$.  
\end{enumerate}
Since only a fraction, $4 \alpha$, of the force is redistributed in each 
relaxation event (toppling), the model is nonconservative for 
$\alpha<1/4$. In the following we concentrate on this case, i.e.  
$ 0< \alpha <1/4$.

To completely define the model we need to specify the boundary conditions.
Boundaries are believed to play a crucial role for the observation of
critical behaviour in the OFC model. It has been suggested that they act 
as inhomogeneities which frustrate the natural tendency of the model to
order into a periodic state~\cite{grass2,socolar}.
Indeed, for sufficiently small values of the conservation parameter 
$\alpha$ ($\alpha < \alpha _c \simeq 0.18$), a system with periodic boundary 
conditions quickly reaches an exactly periodic state with only earthquakes 
of size one. For larger values of $\alpha$ the situation is slightly more 
complicated with multiple topplings involved in a single avalanche, but 
the avalanches are still localised and criticality is not 
observed~\cite{grass2}.
A system with open boundaries is prevented from reaching a periodic state 
because boundary sites have fewer neighbours and therefore cycle at a 
different frequency from bulk sites.  Middleton and Tang suggested that the 
inhomogeneity created by the boundaries propagates into the bulk of the 
system, developing, in this way, long range spatial correlations.  They named 
this mechanism ``marginal synchronisation'' or phase locking~\cite{middleton}. 
In accordance with previous studies, therefore, we consider open boundary 
conditions. If a boundary (corner) site topples, an extra amount $\alpha F_i$ 
($2 \alpha F_i$) is simply lost by the system.

\section{Results: probability distributions}

\begin{figure}[hb]
\narrowtext
\epsfxsize=2.5in
\centerline{\epsffile{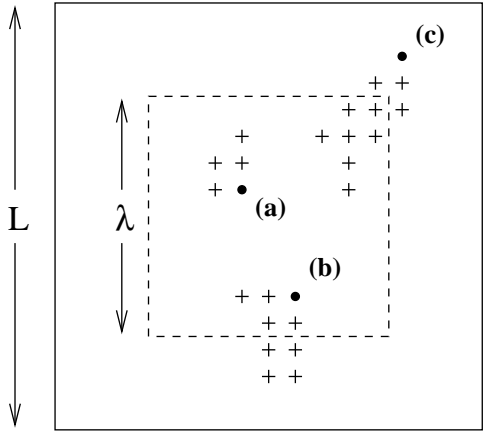}}
\protect\vspace{0.2cm}
\caption[1]{\label{scheme}  
Schematic representation of different types of avalanches.  
The continuous line represents the lattice of size $L$, the dashed line 
the subsystem of linear extent $\lambda$. Triggering sites are denoted 
with a full circle, toppling sites with a cross. Avalanche (a) contributes 
to the distribution $P_{conf}(\lambda,L,s)$; 
(a) and (b) to $P_<(\lambda,L,s)$; (c) to $P_>(\lambda,L,s)$.
}
\end{figure}

After a sufficiently long transient time, the system reaches a stationary 
state. Several statistical properties can be used to characterise this state.
Most previous studies of the OFC model have focused on the behaviour of the 
probability distribution of earthquake sizes, $P_L (s)$, where $L$ is the
size of the system and $s$ is the total number of topplings events during an 
earthquake~\cite{ofc,lisepac,grass2,socolar,middleton,kertesz,corral,ceva}. 
We choose instead to analyse the behaviour of different distributions for 
avalanches sizes, which distinguish between earthquakes according to the 
region of the lattice involved (e.g. bulk or boundary) and 
the coordinates of the triggering site (see figure~\ref{scheme}). This 
investigation is particularly relevant for the OFC model in view of the 
strong inhomogeneity in the spatial distribution of 
avalanches~\cite{grass2,middleton,ceva}. According to ref.~\cite{grass2}, 
for example, large avalanches are localised near the boundary (at least for 
$\alpha < \alpha _c$). As a minor technical remark, we point out that we 
exclude from our data avalanches which involve only one site ($s=1$) as they 
appear to behave according to their own statistics~\cite{grass2}. As we are 
mainly interested in asymptotically large earthquakes, this does not alter our 
conclusions.   

\begin{figure}[hb]
\narrowtext
\epsfxsize=4.in
\centerline{\epsffile{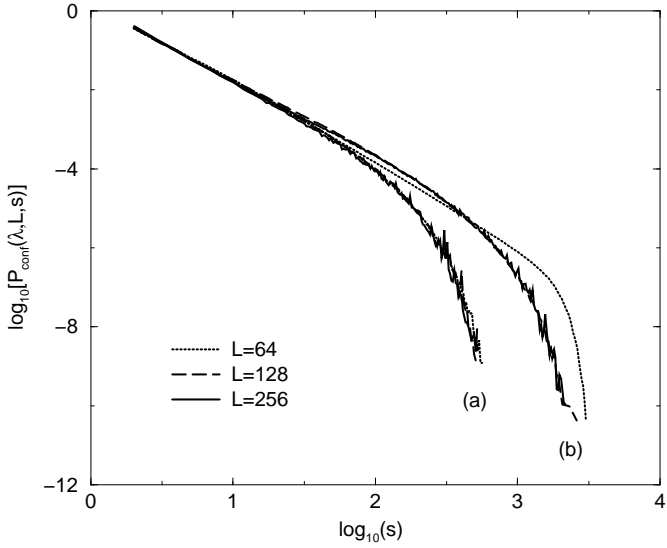}}
\caption[1]{\label{fig_2}  
Probability distribution $P_{conf}(\lambda,L,s)$ for $\alpha=0.18$ and 
(a) $\lambda=32$ and (b)  $\lambda=64$.
}
\end{figure}

We consider a subsystem of linear extent $\lambda$ centred in a system 
of size $L$. The first distribution we introduce, $P_{conf}(\lambda,L,s)$, 
is the normalised distribution of earthquake sizes restricted to earthquakes 
which are confined entirely within the subsystem (e.g. avalanche (a) in 
fig.~\ref{scheme}). The model is driven according to its usual dynamics 
but only those particular earthquakes are counted.  According to our 
definition, the case $\lambda=L$ corresponds to the distribution of
avalanches which do not reach the boundary of the system. 
As shown in fig.~\ref{fig_2}, the distribution  $P_{conf}(\lambda,L,s)$ 
becomes independent of $L$, if $L$ is considerably larger then  $\lambda$
(approximately $L \ge 2 \lambda$). When $L$ approaches $\lambda$, this is no 
longer the case and the cutoff in the distribution is pushed to larger sizes.
Although we have shown in Fig.~\ref{fig_2} only the distributions for 
$\alpha=0.18$ and $\lambda=32,64$, analogous considerations apply to 
different values of $\alpha$ and for different sizes, $\lambda$. Since for
a generic $L$,  $P_{conf}(\lambda,L,s) \neq P_{conf}(\lambda,\lambda,s)$, 
a small portion of a large system is substantially different from a finite 
system of the same size, contrary to what happens in equilibrium critical 
phenomena. A similar observation was made in ref.~\cite{schenk} for the DS 
forest fire model. 
In the following, we denote with $P_{conf}(\lambda,s)$ the distribution 
$P_{conf}(\lambda,L,s)$ in the limit where the distribution does not appear 
to depend on $L$.
In order to determine numerically these distributions, for each value of 
$\lambda$ we have simulated (for at least $2\cdot 10^9$ earthquakes) a system 
of size $L=2 \lambda$. The dependence on $L$ of $P_{conf}$ can in this case be 
safely neglected. With this choice, the accuracy of the measures and  the 
range of scales investigated are optimised, within our computational limits.

\begin{figure}[hb]
\narrowtext
\epsfxsize=4.in
\centerline{\epsffile{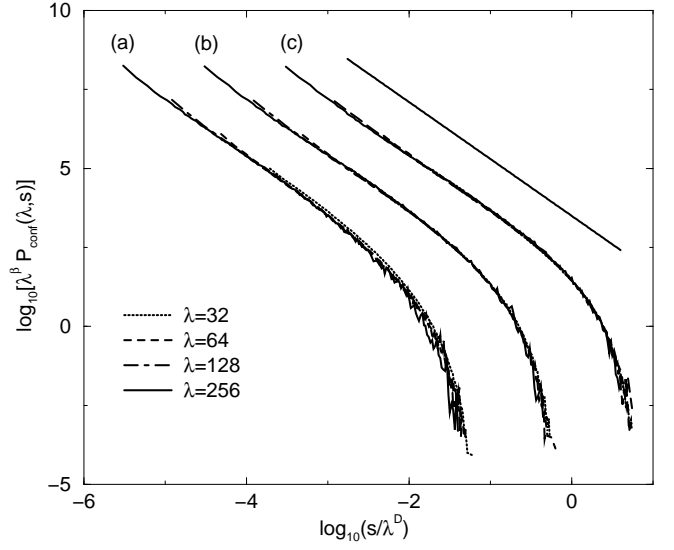}}
\caption[1]{\label{fig_3}
Finite-size scaling plots of $P_{conf}(\lambda,s)$ (with the subsystem 
placed at the centre) for (a) $\alpha=0.15$,
(b) $\alpha=0.18$ and (c) $\alpha=0.21$. The critical exponents are 
$\beta=3.6$ and $D=2$; the slope of the straight line is $\tau =1.8$. 
For visual clarity, curves (a) and (c) have been shifted along the 
$x$-axis, $x \rightarrow x-1$ and  $x \rightarrow x+1$ respectively. 
}
\end{figure}

In Fig.~\ref{fig_3} we report a FSS collapse of $P_{conf}(\lambda,s)$ for 
different values of $\alpha$ . Contrary to the entire distribution of 
earthquake sizes, $P_L(s)$, we observe that $P_{conf}(\lambda,s)$ satisfies 
the FSS hypothesis, i.e. 
$P_{conf}(\lambda,s)\simeq \lambda^{-\beta}f(s/\lambda^D)$, with universal 
critical coefficients. The curve corresponding to $\alpha=0.15$ and 
$\lambda=256$ shows some noisy behaviour, due to the difficulties in
collecting good statistics in this case. Indeed by decreasing $\alpha$, the 
relative fraction of earthquakes in the bulk of the system (with size $s>1$) 
diminishes. Nonetheless, there is no evident  sign that FSS is violated in 
this case. The critical exponents used in the fit of Fig.~\ref{fig_3} are 
$\beta=3.6$ and $D=2$, independent of the dissipation parameter $\alpha$. 
The value of the histogram exponent $\tau= \beta/D \simeq 1.8$ we obtain is 
the same as that found for $P_L(s)$ \cite{lisepac}.  
In addition, the value of $D$ we find corresponds to the largest possible
value for the entire distribution ($D_{max}$ in ref.~\cite{lisepac}), as it 
can be shown that non-conservation requires $D \le 2$ \cite{klein}.

\begin{figure}[hb]
\narrowtext
\epsfxsize=4.in
\centerline{\epsffile{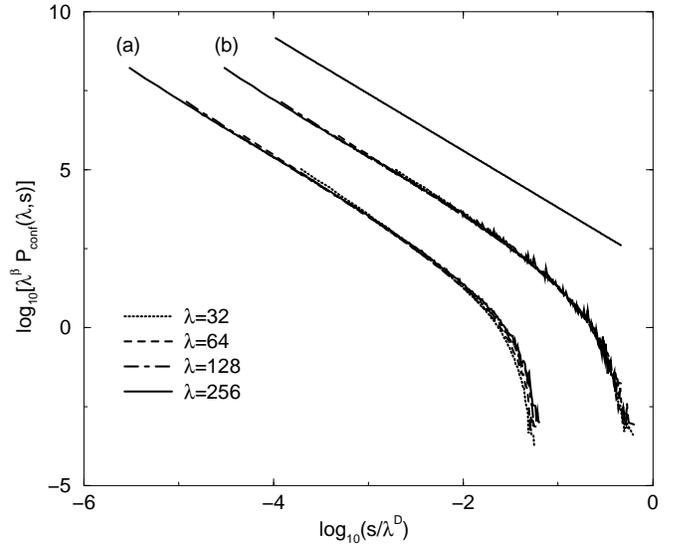}}
\caption[1]{\label{fig_4}
Finite-size scaling plots of $P_{conf}$ for the subsystem placed
(a) on a corner and (b) on a boundary of the system ($\alpha=0.18$). 
The critical exponents are $\beta=3.6$ and $D=2$; the slope of the straight 
line is $\tau =1.8$. Curve (a) has been shifted, $x \rightarrow x-1$. 
}
\end{figure}

The scaling behaviour of $P_{conf}$ appear to be reasonably robust with respect
to translating the subsystem within the entire system; in fig.~\ref{fig_4} we
report a FSS plot for the subsystem placed on a boundary and on a corner of 
the system for  $\alpha=0.18$. While the FSS collapse for the subsystem placed 
on the boundary is rather good, some deviations from FSS are observed in the 
cut-off region for the case of the subsystem placed in the corner. We believe
this behaviour can be ascribed to the enhanced boundary effects in the latter
case (two side of the subsystem are boundary sides instead of only one) and 
would disappear if larger (sub)systems could be studied.
This picture is confirmed by choosing different $\alpha$ values: for the 
subsystem in the corner, deviations from FSS are more pronounced for 
$\alpha=0.21$ and are absent for $\alpha =0.15$. For  the subsystem on the 
boundary, instead, the quality of FSS collapse is rather convincing in all 
cases.

\begin{figure}[hb]
\narrowtext
\epsfxsize=4.5in
\centerline{\epsffile{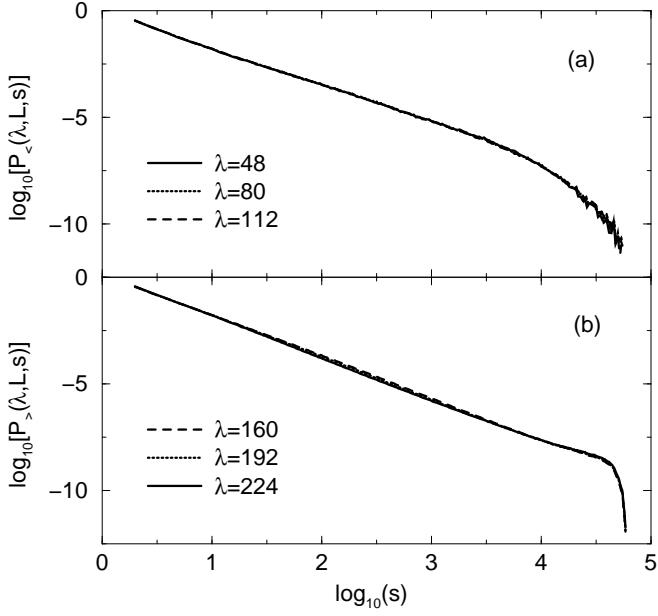}}
\caption[5]{\label{fig_5}
Probability distributions (a) $P_<(\lambda,L,s)$  and (b) $P_>(\lambda,L,s)$
for $\alpha=0.18$, for $L=256$ and for various $\lambda$.} 
\end{figure}

We introduce next the distributions $P_<(\lambda,L,s)$ and $P_>(\lambda,L,s)$.
These are the normalised distribution of earthquakes which start respectively
within ($P_<$) and outside ($P_>$) the subsystem of size $\lambda$,
irrespective of whether they stay in or go out of the subsystem (see 
fig.~\ref{scheme}). 
The only difference between these two distributions, $P_<$ and $P_>$, is the
location of the site that triggers the avalanche. 
We observe numerically that the distributions $P_<(\lambda,L,s)$ and 
$P_>(\lambda,L,s)$ become independent of $\lambda$ respectively in the limit 
$\lambda \ll L$ and $\lambda \simeq L$. As an example we report in 
Fig.~\ref{fig_5} the behaviour of $P_<(\lambda,L,s)$ and $P_>(\lambda,L,s)$
for $\alpha =0.18$, $L=256$ and for various $\lambda$. 
For simplicity, in the following we denote with $P_<(L,s)$ and $P_>(L,s)$ the 
distributions in the limit where they do not depend on $\lambda$. 

\begin{figure}[hb]
\narrowtext
\epsfxsize=4.5in
\centerline{\epsffile{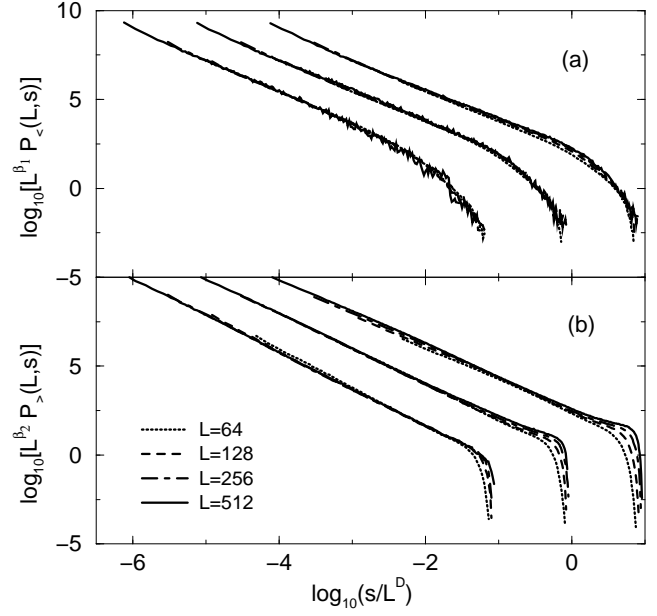}}
\caption[5]{\label{fig_6}
Finite-size scaling plots of (a) $P_<(L,s)$ and (b) $P_>(L,s)$. Different
sets of curves corresponds, from left to right, to $\alpha=0.15$ (shifted 
by $x \rightarrow x-1$), $\alpha=0.18$ and $\alpha=0.21$ (shifted by 
$x \rightarrow x+1$). The exponents are  $\beta _1=3.6$,  $\beta _2=3.9$ 
and $D=2$.}
\end{figure}

We consider therefore two centred subsystems of linear extent 
$\lambda_2>\lambda_1$, such that the above conditions are satisfied.
More specifically, we choose $\lambda_1 = \frac{3}{16} L$ and 
$\lambda_2 = \frac{7}{8} L$. In this case $P_>$ corresponds to the subset 
of earthquakes which are initiated in some ``boundary'' region and $P_<$ 
corresponds to the subset of earthquakes which are initiated within some 
``bulk'' region.
In fig.~\ref{fig_6} we report a FSS scaling plot both for  $P_>$ and
$P_<$. In this figure we choose $D=2$ as the maximum allowed value.  
It is clear that the boundary distribution $P_>$  cannot be collapsed 
according to the FSS ansatz. In fact it develops a sharper and sharper 
cutoff which changes shape and which has an excess of large events (the
cutoff moves towards right for increasing $L$).
The bulk distribution $P_<$ instead does not develop a noticeably sharper 
cutoff and does not appear to change its shape. It may possibly
be collapsed according to the FSS ansatz. Consistent with the
results for $P_{conf}$ and for $P_L$, the power law exponent for  $P_<$
is $\tau= \beta _1/D \simeq 1.8$.

The above numerical analysis indicates that the large events which
violate FSS are triggered by sites in a boundary region. Indeed the 
behaviour of the cutoff for the collapsed probability distributions 
$P_>$ and $P_L$ is very similar (see Fig.~1 in ref.~\cite{lisepac}). 
Although large earthquakes are focussed mainly toward the boundary, 
as suggested in ref.~\cite{grass2}, they occur also in the bulk of the
system, even for low values of $\alpha$, as can be deduced from 
Fig.~\ref{fig_3} and Fig.~\ref{fig_6}. Moreover, we do not observe any
significant qualitative change in the behaviour of the system around 
$\alpha = \alpha _c \simeq 0.18$, as claimed in  ref.~\cite{grass2}.

\section{Discussion and conclusions}

Similarly to other SOC systems~\cite{stella,schenk,pastor_1,pastor_2}, the 
nonconservative OFC model shows relevant deviations from simple 
FSS~\cite{lisepac}.
In this paper we have investigated the origin of this phenomenon, finding 
that FSS in the OFC model is violated because of large, system wide 
earthquakes. In fact, we have found that earthquakes localised within a given 
subsystem do obey ordinary FSS, with universal critical exponents,
independently of whether the subsystem is placed at the centre or on a 
boundary of the system. The value of the power law exponent, 
$\tau \simeq 1.8$, for the ``confined'' distribution agrees with the one for 
the entire distribution. We have shown, moreover, that the probability 
distribution for earthquakes initiated in a boundary region do not obey
FSS, because of an ``excess'' of large events. This would result in an 
apparent exponent $D>2$ which is not allowed in the nonconservative case.
On the other hand, the probability distribution for earthquakes starting
in the bulk of the system is compatible with a FSS hypothesis. In particular, 
the critical exponent is $D=2$ in this case, indicating that large 
earthquakes responsible for breaking finite size scaling are initiated 
predominantly near the boundary of the system.

Self-organised criticality in the OFC model has been ascribed to a mechanism 
of ``marginal synchronisation''~\cite{middleton}. A  system with open 
boundaries becomes almost synchronised by an invasion process where spatial 
correlations develop from the boundaries.
It was suggested that sites close to the boundaries start to organise
themselves first, building up long range correlations. The critical region
grows with time, until, in the stationary state, it invades the whole
lattice~. Our findings on the large events occurring at the 
boundary seem to indicate that the effect of synchronisation is stronger for 
boundary sites than for  bulk sites. This view is supported also by the 
``on screen'' observation that large earthquakes tend to be triggered 
repetitively by the same sites over a long time scale (a result which seems 
to be confirmed also by the study in ref.~\cite{ceva}).

\medskip

S.L. acknowledges financial support from EPSRC (UK),  
Grant No. GR/M10823/01.

\end{multicols}
\end{document}